\documentclass[aip,apl,superscriptaddress,amsmath,amssymb]{revtex4-1}
\usepackage{graphicx}% Include figure files
\usepackage{dcolumn}% Align table columns on decimal point
\usepackage{bm}% bold math
\usepackage{textcomp}
\usepackage[utf8]{inputenc}

\newcommand{\He}{$^{3}$He }
\newcommand{\Ni}{Ni$_{3+x}$Al$_{1-x}$ }

\begin{document}

%\preprint{}

%\title{Visualization of the suppression of the ordering temperature of the weak itinerant ferromagnet Ni$_3$Al under hydrostatic pressure by neutron depolarization imaging} 

\title{Neutron depolarization imaging of the hydrostatic pressure dependence of inhomogeneous ferromagnets} 



\author{M. Schulz}
%\email[]{michael.schulz@frm2.tum.de}
\affiliation{Physik-Department, Technische Universit\"{a}t M\"{u}nchen, James-Franck-Stra\ss e, 85748 Garching, Germany}
\affiliation{Heinz Maier-Leibnitz Zentrum (MLZ), Technische Universit\"{a}t M\"{u}nchen, Lichtenbergstr. 1, 85748 Garching, Germany}

\author{A. Neubauer}
%\thanks{}
\affiliation{Physik-Department, Technische Universit\"{a}t M\"{u}nchen, James-Franck-Stra\ss e, 85748 Garching, Germany}

\author{P. B\"{o}ni}
\affiliation{Physik-Department, Technische Universit\"{a}t M\"{u}nchen, James-Franck-Stra\ss e, 85748 Garching, Germany}

\author{C. Pfleiderer}
\affiliation{Physik-Department, Technische Universit\"{a}t M\"{u}nchen, James-Franck-Stra\ss e, 85748 Garching, Germany}

\date{\today}

\begin{abstract}
The investigation of fragile and potentially inhomogeneous forms of ferromagnetic order under extreme conditions, such as low temperatures and high pressures, is of central interest for areas such as geophysics, correlated electron systems, as well as the optimization of materials synthesis for applications where particular material properties are required. We report neutron depolarization imaging measurements on the weak ferromagnet Ni$_3$Al under pressures up to 10\,kbar using a Cu:Be clamp cell. Using a polychromatic neutron beam with wavelengths $\lambda\geq$ 4\,\AA\, in combination with \He neutron spin filter cells as polarizer and analyzer we were able to track differences of the pressure response in inhomogeneous samples by virtue of high resolution neutron depolarization imaging. This provides spatially resolved and non-destructive access to the pressure dependence of the magnetic properties of inhomogeneous ferromagnetic materials. 
\end{abstract}


\maketitle %\maketitle must follow title, authors, abstract and \pacs

%***********************************************************************************************
The non-destructive and spatially resolved investigation of inhomogeneous ferromagnetic properties of materials under pressure has for a long time been a great challenge in solid state research. Progress on this question concerns a wide range of scientifically interesting fields like geophysical investigations, fundamental research in correlated electron systems as well as applied research, such as material synthesis and its optimization for tailored applications. For example, investigation of the distribution and development of magnetic properties under pressure in natural minerals like magnetite is of paramount importance in geophysical research to obtain a detailed understanding of the composition and development of the earth's interior\cite{Gilder2013Fe_under_pressure}. In this context it can be crucial to use methods which allow for non-destructive and in-situ probing of the minerals. Similar prerequisites can also be found in applied materials science where non-destructive and spatially resolved probing of magnetic properties in volume or thin film material can be essential for the synthesis and tailoring of (in)homogeneous magnetic properties. Lastly, the effect of pressure on the electronic density of state at the Fermi surface and the residual effects on the superconducting or magnetic behavior is of elementary importance in the physics of correlated electron systems.

In recent years radiography with polarized neutrons is increasingly being recognized as a powerful method for the investigation of magnetic field distributions, most notably of internal magnetic fields in bulk materials and trapped flux in superconductors\cite{kardjilov2008tdi, piegsa2009qrm, Treimer2012APL_Partial_Meissner_Effect, Treimer2012PRB_Flux_Trapping}. As a special application of this technique, neutron depolarization imaging\cite{mschulz_diss} is based on the measurement of the change of the neutron beam polarization after transmission of a ferromagnetic sample. In ferromagnetic materials the interaction of the neutron's nuclear magnetic moment $\mu$ with the magnetic field distributions in the domains leads to a depolarization of the neutron beam, resulting in an imaging contrast. Contrary to this, paramagnetic and antiferromagnetic materials do not change the beam polarization. Yielding a spatially resolved image of the ferromagnetic regions in a bulk material this technique has the advantage of being fast and non-destructive. Particularly in doped ferromagnetic or superconducting samples with a tendency to show inhomogeneous magnetic behavior over the sample size, the magnetic properties at different positions can be directly investigated and compared without the tedious and destructive preparation of single pieces for bulk measurements. In this context it has already been shown that even a 3D tomographic reconstruction of the magnetic properties is possible\cite{kardjilov2008tdi, mschulz_PNSXM2009_depol_tomo}. 

Besides polarizers providing a high neutron beam polarization while not affecting the beam collimation\cite{mschulz_ICNS2009} a high flux neutron beam leading to good statistics is a prerequisite for neutron depolarization imaging.  We have employed a polychromatic neutron beam with polarization analysis using \He, which allows us to perform high resolution magnetic imaging with good counting statistics on large samples and under high pressure despite the additional material of the pressure cell in the beam. In this paper we present neutron depolarization imaging measurements under pressure, using a thinned Cu:Be clamp cell with reduced neutron absorption. With this setup a spatial resolution as good as 0.3\,mm may be achieved. 

From previous investigations of the resistivity of the weak itinerant ferromagnet Ni$_3$Al a strong pressure dependence of the magnetic properties is well established. Application of a pressure of $\sim$80 kbar reduces $T_C$ from $\sim$45\,K towards zero, giving way to a ferromagnetic quantum phase transition\cite{niklowitz2006Ni3Al}. A deviation from the stoichiometric composition in \Ni by increasing the Ni concentration $x$ leads to an increase of the ordering temperature\cite{Dhar1989Ni3Al}. It is known from metallurgic investigations that the growth of Ni$_3$Al is very difficult due to a peritectic point in a very narrow temperature range\cite{bremer1988Ni3Al}. However, due to the strong dependence of the ordering temperature $T_C$ on the composition any metallurgic inhomogeneity will consequently lead to a variation of the magnetic properties of the sample, generating great uncertainties in studies of the behaviour at the quantum phase transition. In this context we have investigated the pressure dependence of the ferromagnetic transition temperature, ${dT_{C}}/{dp}$, in a \Ni ($x \approx 0.02$) sample to explore the potential of the neutron depolarization imaging technique for this and related fields. 

\begin{figure}[]
 \includegraphics[width=0.48\textwidth]{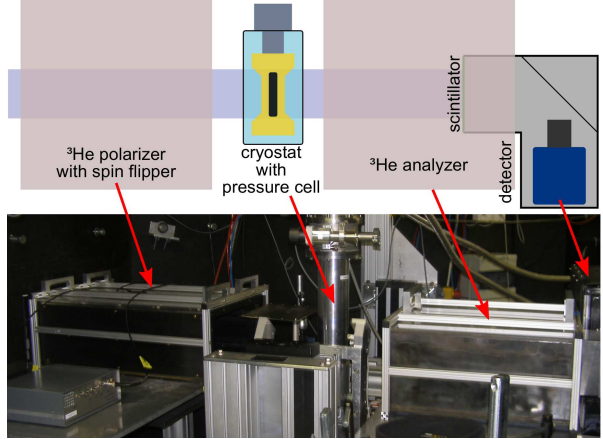}
 \caption{Schematic setup for polarized neutron radiography at the ANTARES beam line at MLZ (top). A \He polarizer with an integrated adiabatic fast passage (AFP) spin flipper, a closed cycle cryostat with a base temperature of 5\,K, a \He analyzer, a neutron sensitive LiF:ZnS scintillator and a position sensitive CCD detector. Photo of the setup (bottom)
}
 \label{setup}
\end{figure}

The neutron depolarization imaging experiments have been performed at the ANTARES beam line at MLZ, Garching\cite{schillinger2004dnr, calzada2005_ANTARES,schulz2015ANTARES_JLSRF}.  The set-up shown in Fig. \ref{setup} consists of a \He polarizer with integrated spin flipper\cite{Babcock2007_3He_flipperizer}, a Helium free closed cycle cryostat with a base temperature of 5\,K, a \He analyzer. Furthermore, a neutron sensitive scintillator is placed behind the analyzer and a CCD detector is used to record the resulting image. The polarization between the two \He spin filters was maintained by vertical guide fields with a magnitude $B \approx 2$\,mT. The \He cells were prepared at the HELIOS facility\cite{Hutanu20073He} at MLZ and filled with polarized \He gas at a pressure of 1.6\,bar. The thickness of the cells in beam direction was 5\,cm and the maximum cross section of the polarized neutron beam is $10 \times 10\,{\rm cm^2}$ which is given by the size of the \He cells. The transmission of a single cell for the unpolarized, polychromatic beam was $29\%$.

The minimum distance between sample and detector is restricted to $d \approx 300$\,mm by the \He cell that must be placed in a bulky magneto-static cavity, which limits the geometrically achievable imaging resolution to  approximately 0.3\,mm at a collimation ratio of $L/D = 800$. The magneto-static cavity is required to produce a magnetic field $B \approx 2$\,mT and a homogeneity of $\Delta B/B<10^{-4}$ at the position of  the \He container to maintain a high relaxation time of the polarized \He gas of $\tau \approx$ 200\,h. The slow depolarization of the \He gas with time was corrected for in the data evaluation. One of the magneto-static cavities was equipped with a radiofrequency coil setup that allowed to flip the polarization of the \He gas and consequently also the beam polarization\cite{Babcock2007_3He_flipperizer}. The flipping ratio $R\approx 10$ obtained with the \He cells was determined to be $R \approx 10$ for freshly polarized gas resulting in an average beam polarization of $P_0=82\%$ for the polychromatic beam. 

For a random orientation of the ferromagnetic domains in the sample and under the assumption of a small net rotation of the neutron spin within a single domain the change of the beam polarization after transmission of the sample may be written as\cite{halpern1941depol} 
\begin{equation}
P=P_0\exp{\left(-\frac{1}{3}\gamma^2B^2\frac{d\delta}{v^2}\right)},
\label{Eq_Halpern}
\end{equation}
 where $P_0$ is the initial beam polarization, $\gamma=1.83\cdot10^8\mathrm{s}^{-1}\mathrm{T}^{-1}$ is the gyromagnetic ratio of the neutron, $B$ is the magnitude of magnetic field within a single domain, $d$ is the sample thickness, $\delta$ is the average domain size and $v$ is the neutron velocity. This formula relates the magnetic properties of the sample with the beam polarization.

For weak itinerant ferromagnets in a Ginzburg-Landau approach the magnetization $M(T)$ is typically assumed to take the form\cite{lonzarich1985fluctuations} $M^2(T)=M_{0}^2(1-(T/T_C)^2)$. Using this expression the magnetic field in the domains may be written as $B(T)=\mu_0 M(T)$, which introduces a temperature dependence in eq. \ref{Eq_Halpern}.

These measurements were performed using a polychromatic neutron beam with wavelengths $\lambda \geq$ 4\,\AA\ which was obtained by inserting a polycrystalline Be filter into the neutron beam\cite{Lorenz2007_Filterrrad}. Using a polychromatic beam increases the neutron flux and consequently the counting statistics of the measurement compared to a monochromatic beam\cite{mschulz2009mono} by approximately a factor of 30. While the dependence of the transmitted beam polarization on the neutron velocity $v$ in eq. \ref{Eq_Halpern} hinders the exact determination of the average domain size and magnetization of the sample for a polychromatic beam the onset of depolarization (and thus $T_C$) can still be determined.

\begin{figure}[]
 \includegraphics[width=0.48\textwidth]{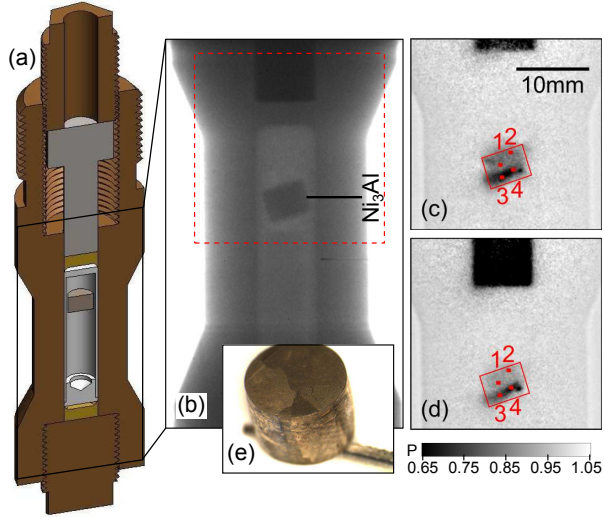}
 \caption{Technical drawing of the thinned Cu:Be pressure cell (a). The sample is placed inside a Teflon cup filled with Fluorinert as a pressure medium. Neutron absorption radiograph (b) of the pressure cell containing a \Ni sample. Polarization of the neutron beam after transmission of the sample at $T=82$\,K within the dashed red region shown in (b) for a pressure of $p=0.2$\,kbar (c) and 10\,kbar (d). The red boxes (1 - 4) mark the regions over which the temperature dependence of the beam polarization is plotted in Fig. \ref{pol}. Spatial variations in the neutron depolarization can be observed over the sample cross section due to sample inhomogeneities. Photo of the sample (e)}
 \label{DZ}
\end{figure}

Hydrostatic pressure was applied by means of a Cu:Be clamp cell with a reduced radius around the sample position to minimize neutron absorption (see Fig. \ref{DZ}(a)). A Fluorinert FC72:FC84 mixture at a 1:1 volume ratio was used as pressure transmitter. The sample space within the Teflon cup in the pressure cell had a diameter of 7\,mm and a maximum height of 3\,cm. The wall thickness of the Cu:Be alloy around the sample space was reduced to 7\,mm. The applied pressure was calculated from the applied force used to load the cell.

A polycrystalline \Ni rod with $x \sim 0.02$ was prepared by float zoning in an optical image furnace\cite{Neubauer2011ImageFurnace}. A disc with a height of 5\,mm and a diameter of 7\,mm was cut from the rod (photo shown in Fig. \ref{DZ}(e). Inhomogeneities within the disc have been found both in previous neutron radiography\cite{mschulz_diss} and susceptibility measurements, which showed that the sample mainly consists of three grains with different properties. An average ordering temperature of $T_C \approx 80$\,K with a variation of approx. $\pm10$\,K was found. The disc was placed in the clamp cell together with a paramagnetic Fe$_2$VAl sample (used for a different study and barely visible in the radiography in Fig. \ref{DZ}(b)) using aluminum foil as a spacer. 

Neutron depolarization imaging measurements at different temperatures were performed at two different pressures $p_1=0.2$\,kbar and $p_2=10$\,kbar. The low pressure measurement was performed at a slight overpressure of 0.2\,kbar above ambient pressure (pressure losses permitting) in order to maintain the orientation of the sample inside the pressure cell.  Data were recorded for temperatures at 70\,K and between 75\,K and $\approx 90$\,K in 1\,K steps. For all temperature steps a set of 30 spin-up ($I_\uparrow$) and spin-down ($I_\downarrow$) images with an exposure time of 60\,s per image were taken. Consequently the total measurement time at each temperature summed up to 60 minutes. The overall measurement time for this experiment, including the time to stabilize the sample temperature at each temperature step was approximately one day. The polarization was calculated from the measured images as $P=1/P_0(I_\uparrow-I_\downarrow)/(I_\uparrow+I_\downarrow)$, where $P_0$ is the polarization of the beam with no sample in place.

Fig. \ref{DZ}(b) - (d) show a typical set of images as obtained for each temperature step (here for $T=82$\,K).  A standard neutron radiography image is shown in (b) in which the contrast arises due to the nonmagnetic interaction of the neutrons with the sample. The thinned shape of the pressure cell is clearly visible as well as the dark pressure piston on the top side, which is made from tungsten carbide (WC). Furthermore, the position of the \Ni sample located in the topmost third of the brighter pressure volume is indicated by a black line in the image. 

Fig. \ref{DZ}(c) and (d) show the neutron beam polarization $P$ after transmission of the sample within the dashed red region indicated in (b) at $p=0.2$\,kbar and 10\,kbar, respectively. The polarization is coded between black ($P=0.65$) and white ($P=1.05$, see color bar). This scaling was chosen to visualize the small depolarization of the beam observed in the experiment. One may clearly observe the lower position of the slightly ferromagnetic - and thus depolarizing - pressure piston at higher pressure in Fig. \ref{DZ}(d). At a first glimpse one may already notice the reduction of the darkened area at higher pressure, i.e. the increase of the neutron beam polarization with increasing pressure. 
%Although the pressure cell is non-magnetic, it is still slightly visible in the polarization images (c) and (d). This effect is most likely caused by a small amount of stray light in the detector box which is reflected back by the mirror to the scintillator. Due to the very low count rates behind the massive pressure cell even a small amount of scattered light leads to a significant change of the detected intensity. However, this effect only leads to a constant offset in the measured beam polarization of approximately 1\% and can therefore be neglected in the context of this study.%
Different gray shading within the sample arises due to spatially inhomogeneous magnetic properties of the sample.

\begin{figure}[]
\includegraphics[width=0.48\textwidth]{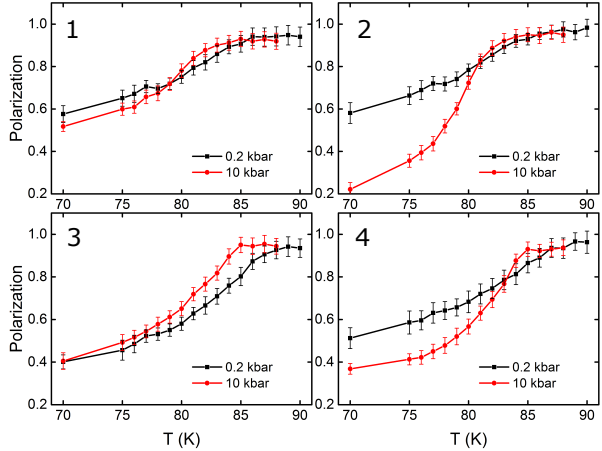}
\caption{Temperature dependence of the neutron beam depolarization by a \Ni sample averaged over the red rectangles 1 - 4 shown in Fig. \ref{DZ}. Application of a hydrostatic pressure of 10\,kbar leads to a reduction of the ordering temperature of approximately 3\,K in regions 3 and 4, while $T_C$ does not change significantly in regions 1 and 2. Due to a residual magnetization of the sample, which sets in at around 215\,K the beam polarization does not reach 100\%.}
 \label{pol}
\end{figure}

Fig. \ref{pol} shows the temperature dependence of the neutron beam polarization averaged over four different areas 1 - 4 of the sample marked by red rectangles in Fig. \ref{DZ} (c) and (d) which are representative for the variation of the behavior of $P(T)$ observed over the sample. 
For each area the high pressure data is shown in red, while the low pressure data is displayed in black. The error bars show the standard deviation of the polarization value in the regions, which depends both on the counting statistics and the homogeneity of $P$ within the region. 
A comparison of the high and low pressure data for regions 1 and 2 shows no significant change of the onset of neutron depolarization, which coincides with the onset of ferromagnetic ordering at these positions of the sample. In contrast, the data for regions 3 and 4 clearly show a decrease of the ordering temperature under an applied pressure of 10\,kbar, which is estimated to lie approximately 3\,K below the value obtained for low pressure (87\,K). 
This gives a pressure dependence of the ordering temperature of $\partial T_{\mathrm{C}}/{\partial p}\approx-0.3$K/kbar based on the assumption of a linear decrease of $T_C$ with pressure. This value is lower than the value of -0.5\,K/kbar found by Niklowitz et al.\cite{niklowitz2006Ni3Al} for a stoichiometric sample. These findings clearly suggest that our sample has metallurgic inhomogeneities which lead to a variation of the magnetic properties over the sample volume via the concentration dependence $T_C(x)$ in \Ni\cite{Dhar1989Ni3Al}. Furthermore, we observe a constant depolarization of the beam by the sample even at higher temperatures. This is attributed to a residual magnetization of the sample, which was found to set in at $\sim$215\,K as inferred from ambient pressure susceptibility measurements on this sample and might be caused by a minor contamination of the sample with a Ni-rich phase. 

The temperature dependence of the beam polarization below the ferromagnetic transition in all regions shows a steeper decrease with applied pressure than at ambient pressure. Furthermore, in regions 2 and 4 at 70\,K the beam polarization under pressure is significantly below the polarization observed at ambient pressure. This behavior is intuitively unexpected since the application of pressure typically leads to a decrease of the magnetic moments and therefore also of the magnetic field $B$ inside the domains. According to eq. \ref{Eq_Halpern} this should consequently cause the beam polarization to increase at a certain temperature compared to the ambient pressure value. Possible reasons for the reduction of the beam polarization with pressure are an expansion of the domain size $\delta$ in the sample or a reorientation of the domains due to uniaxial pressure components introduced by the freezing of the pressure medium in combination with the magnetostriction of the material\cite{katsuhiko1985Ni3Al_Magnetostriction}. Another important effect may be strain within the sample due to small differences of the compressibilities that originate in differences of composition. All effects could lead to a reduction of the beam polarization. To clarify this behaviour observed, further experiments - preferably with several homogeneous samples - will be necessary.

Our experiments on a \Ni sample in a Cu:Be pressure cell have demonstrated the possibility to investigate the pressure dependence of ferromagnetic phase transitions by means of neutron depolarization imaging. We obtained spatially resolved information on the pressure dependence of the magnetic properties of a bulk material in a single measurement and regions within the sample with different magnetic behavior under pressure could be identified. While a monochromatic beam is required for the quantitative investigation of magnetization and domain sizes, the ordering temperature $T_C$ can be determined using a polychromatic beam. By employing a polychromatic neutron beam in combination with polarization analysis we could significantly increase the count rate and therefore improve the counting statistics.  This establishes that neutron depolarization imaging offers the unique possibility of a swift and non-destructive investigation of the magnetic properties in large and inhomogeneous samples, which can be of great interest in many fields of research. To achieve higher pressures or to perform measurements on larger samples, many different types of pressure cells are available, which could be adapted for the use in neutron depolarization imaging experiments. Furthermore, investigations of the development of magnetic domains under hydrostatic as well as uniaxial pressure represent another possibility. 

It has already been shown that a tomographic reconstruction of neutron depolarization imaging data is possible\cite{mschulz_PNSXM2009_depol_tomo} thus providing 3D information on the distribution of ferromagnetic regions in the sample under high pressure. Currently tomographic neutron depolarization experiments are very time consuming due to the limited neutron flux available after monochromatization and polarization of the beam. However, the construction of new high-flux pulsed neutron sources such as the European Spallation Source in Lund, Sweden with intrinsic monochromatization based on a Time-of-Flight approach will offer large gain factors over static reactor based sources. This, in combination with further developments of the neutron depolarization imaging technique by improving polarizing equipment and using focusing neutron optics, will result in much higher flux and better spatial resolution\cite{Kardjilov2010Elliptic_Guide}. Such developments may in the future offer possibilities to enhance the potential of neutron depolarization imaging for the investigation of a wide variety of problems in fields like magnetism, strongly correlated electron systems, superconductivity and geophysics.

\begin{acknowledgments}
The authors wish to thank A. Bauer, P. Jorba and G. G. Lonzarich for helpful discussions and S. Masalovich for providing the polarized \He gas. Financial support from the German Science Foundation (DFG) in the research unit “Quantum phase transitions” (FOR960), and the Augsburg-Munich transregional collaborative research network “From electronic correlations to functionality” (TRR80) is gratefully acknowledged. 
\end{acknowledgments}

\bibliography{depolpaper}

\end{document}